\documentclass{natureprintstyle}
\usepackage{epsfig,caption}
\usepackage{color}
\usepackage{bm}
\usepackage{graphicx}
\usepackage{longtable}
\usepackage{amssymb}
\usepackage{rotating}
\usepackage{latexsym}
\usepackage{hyperref}

\usepackage{graphicx,color}
\usepackage{amsmath} 

\title{A massive, quiescent galaxy at redshift of z=3.717}

\author{
Karl Glazebrook$^{1}$,
Corentin Schreiber$^2$,
Ivo Labb\'e$^2$,
Themiya Nanayakkara$^{1},$ \\
Glenn G. Kacprzak$^{1}$
Pascal A. Oesch$^3$,
Casey Papovich$^4$, \\
Lee R Spitler$^{5,6}$,
Caroline M. S. Straatman$^7$,
Kim-Vy H. Tran$^4$,
Tiantian Yuan$^8$
}

\def\lesssim{\mathrel{\hbox{\rlap{\hbox{\lower4pt\hbox{$\sim$}}}\hbox{$<$}}}}
\def\gtrsim{\mathrel{\hbox{\rlap{\hbox{\lower4pt\hbox{$\sim$}}}\hbox{$>$}}}}

\def\micron{\hbox{$\mu$m}}
\let\la=\lesssim			
\let\ga=\gtrsim

\definecolor{ao}{rgb}{0.0, 0.5, 0.0}

\newcommand\reduline{\bgroup\markoverwith{\textcolor{red}{\rule[-0.5ex]{2pt}{0.4pt}}}\ULon}

\def\ourobject{{ZF-COSMOS-20115}}

\begin{document}

\maketitle

\begin{affiliations}
\item Centre for Astrophysics and Supercomputing, Swinburne
  University of Technology, P.O. Box 218, Hawthorn, VIC 3122,
  Australia
\item Leiden Observatory, Leiden University, PO Box 9513, 2300
RA Leiden, Netherlands.
\item Geneva Observatory, University of Geneva, Ch. des Maillettes 51, 1290 Versoix, Switzerland
\item George P. and Cynthia W. Mitchell Institute for Fundamental Physics and Astronomy, Department of Physics and Astronomy, Texas A \& M University, College Station, TX 77843.
\item Macquarie Research Centre for Astronomy, Astrophysics \& Astrophotonics, Macquarie University, Sydney, NSW 2109, Australia
\item Australian Astronomical Observatory, PO Box 915, North
Ryde, NSW 1670, Australia.
\item Max Planck Institute for Astronomy, K\"onigstuhl 17, 69117 Heidelberg, Germany
\item Research School of Astron.~Astrophys., The Australian National University, Cotter Road, Weston Creek, ACT 2611, Australia.
\end{affiliations}


\begin{abstract}
In the early Universe finding massive galaxies that have stopped forming stars present an observational challenge
as their rest-frame ultraviolet emission is negligible and they can only be reliably identified by extremely deep near-infrared surveys.
These   have revealed the presence of massive, quiescent early-type galaxies\cite{Dunlop96,Cimatti04,McCarthy04,Cimatti08,Gobat2012,Belli2014} appearing in the
universe as early as z$\sim$2, an epoch 3 Gyr after the Big Bang. Their age and formation processes have now been 
explained by an improved
generation of galaxy formation models\cite{Wellons+15,Behroozi+16,Dave+16} where they form rapidly at z$\sim$3--4,  consistent
with the typical masses and ages derived from their observations.
Deeper surveys have now reported evidence for populations of massive, quiescent galaxies at even higher redshifts and earlier times, however the evidence for their existence, and redshift, has relied entirely on coarsely sampled photometry.
These early massive, quiescent galaxies are not predicted by the latest generation
of theoretical models.\cite{Wellons+15,Behroozi+16,Dave+16,MBII}
Here, we report the spectroscopic confirmation of one of these galaxies at redshift z=3.717 with a stellar mass  of 1.7$\times$10$^{11}$ M$_\odot$ whose absorption line spectrum shows no current star-formation and which has a derived age of nearly half the
age of the Universe at this redshift.
The observations demonstrates that the galaxy must have quickly formed the majority of its stars within the first billion years of cosmic history in 
an extreme and short starburst. This ancestral event is 
similar to those  starting to be found by sub-mm wavelength surveys\cite{Blain2002,Capak2011,ALMAz6,Ma2016} pointing
to a possible connection between these two populations. Early formation of such massive systems is likely to require significant 
revisions to our picture of early galaxy assembly.
\end{abstract}


The   massive  galaxy \ourobject\ was selected from a sample of galaxies\cite{Straatman14} identified photometrically as having a spectral break between the H$_{\rm long}$ (1.7\micron) and K$_{\rm s}$ (2.2\micron)  bands. This break, identified as the redshifted  hydrogen Balmer limit in intermediate age stellar populations, together with the full 36 band photometric spectral energy distribution (SED), suggests that these are quiescent at $3.5<z<4.1$, with strongly suppressed star formation, stellar ages of 0.5--1 Gyr and large stellar masses ($\simeq 10^{11}$ M$_\odot$) driven by their K-band brightness. Alternate redshifts and SED solutions were strongly ruled
out by the combination of medium and narrow band photometry.  \ourobject\ is the brightest  of this
sample with K$_{AB}=22.4$ and has a typical SED for this population. It is very compact and red with an effective
radius\cite{Straatman15} of only 0.5 kpc. We obtained a K-band spectrum (Figure~1)  which shows a significant  continuum and clear Balmer absorption lines of H$\beta$, H$\gamma$ and H$\delta$ at $z=3.717$. This is a significantly earlier epoch (1.65 Gyr after the Big Bang) than previous spectroscopy of massive quiescent galaxies.\cite{Dunlop96,Cimatti04,McCarthy04,Cimatti08,Gobat2012,Belli2014} The most distant prior example\cite{Gobat2012} at $z=2.99$
had spectrophotometry that required formation at $z>4$, however the grism spectrum was too low resolution to show absorption lines
to confirm the nature of the stellar population.
Balmer lines are a quintessential feature
of a `post-starburst' type-spectrum because they arise from A-type stars with lifetimes of 200--1000 Myr.
Spectroscopic absorption lines can provide robust age constraints as they identify specific time-sensitive
stellar populations, whereas estimates based on multi-band photometry for red galaxies are very degenerate with the dust attenuation and can be subject to systematic errors between different instruments at different wavelengths\cite{Papovich2006}.
The absence of strong emission lines implies the previous SED-fitting
and mass-estimate were robust.

\begin{figure}
\begin{center}
\includegraphics[width=9cm]{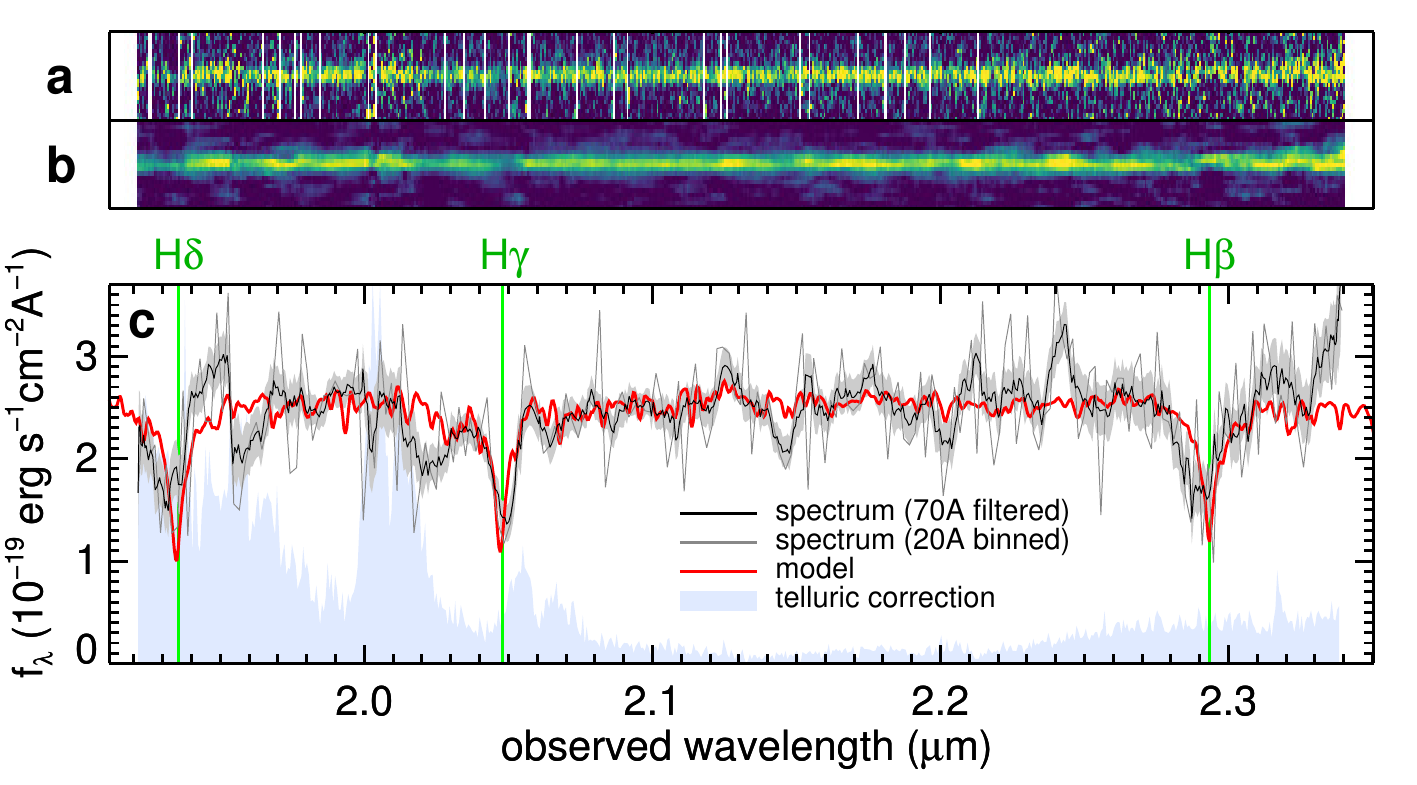}
\end{center}
\label{fig:1Dspec}
\caption*{\small {\bf Figure 1:} Spectrum of \ourobject\ in the near-infrared \textit{K} band. The wavelength axis is the same in all three panels. The galaxy was also observed in \textit{H} band, but continuum was
not detected, which is consistent with the level expected from the photometric break between the \textit{H} and \textit{K} bands. \textit{a}, Original 2D sky-subtracted \textit{K}-band spectrum
from MOSFIRE at its native resolution ($\lambda/\Delta\lambda=3200$). The vertical white lines show
where strong night sky residuals were masked. \textit{b}, Same 2D spectrum optimally smoothed with a boxcar filter to a lower resolution of 70\AA, to enhance visibility of the continuum and broad absorption lines. \textit{c}, The optimally-extracted 1D spectrum, both binned at 20\AA\ resolution and smoothed by a boxcar filter of 70\AA\ to enhance the main absorption features, as in \textit{b}.
Three strong Hydrogen Balmer absorption lines (H$\beta$, H$\gamma$ and H$\delta$) are clearly seen, characteristic
of a post-starburst spectrum.  Balmer emission lines are not seen, confirming the lack of current star-formation.
An example post-starburst template fit is overlaid. The grey band indicates the 1$\sigma$ noise level, and  the regions most affected
by telluric corrections are also indicated by the labelled shading.}
\end{figure}

In Figure~2 we show how the strength of the absorption
lines by themselves strongly constrain the age of the galaxy. We use the integrated equivalent
width of the combined H$\beta\gamma\delta$ absorption lines, which are the most robust measurement we can make from the spectra,
and compare with stellar population models where there is initial period of intense star-formation, followed by a period of quiescence
(full details are given in the Supplementary Section). It takes at least 100--200 Myr  of quiescence to
reach the line strengths needed and the quiescent star-formation rate has to be at least
 a 100$\times$ less than during the formation period. The youngest possible
spectroscopic age is 200--700 Myr and is obtained from near-instantaneous formation ($\la 50$ Myr) and solar abundances in the stellar population. Younger ages, while potentially more compatible with cosmological constraints (as discussed below),
require short formation times and very high early star-formation rates ($>2000$M$_\odot$ yr$^{-1}$).
If star formation is more extended in time, or the  galaxy is metal-poor,  the the implied age of this galaxy would be significantly older (see Figure~2).

With a firmly established spectroscopic redshift the photometry itself is also now much more constraining on the formation history  and
we can rule out degeneracies with dusty SEDs and
contamination from emission lines. We ran a large grid of our star-formation history models (see Supplementary Section) to optimally combine the spectroscopic and
photometric information. This gives a combined constraint on the stellar age of $\simeq 700$ Myr and a much tighter
constraint on the formation time with rapid formation in a period of $\la$ 250 Myr. Full results
and error ranges are given in Table 1, the age uncertainty is only 16\% the age of the Universe at $z=3.717$ which is a significant
improvement on lower redshift results\cite{Dunlop96,Cimatti04,McCarthy04,Cimatti08,Gobat2012,Belli2014} and in particular does not include $z<4$ solutions. The stellar mass we see is relatively unobscured. Since there
are degeneracies between star-formation history parameters the most robust way to present the mass assembly history of
this galaxy is to marginalise across this grid of models to calculate bounds on the stellar mass as a function of cosmic time and redshift (Figure~3). 
Marginalising the rest-frame optical SED modelling also shows that $<$0.1\% of the observed stellar mass formed within the last 200 Myr.
Current star-formation is either negligible ($<4$ M$_\odot$ yr$^{-1}$ from H$\beta$ emission, consistent with the SED), or so extremely obscured that it is not visible at rest-frame 1--2 \micron\ mimicking the purely passive SED.
Longer wavelength data (see Methods) reveals a   faint nearby ALMA source ($\simeq$6 effective radii distant), if part of a more extended system this would
give an obscured star-formation limit of $\lesssim$ 50--200 M$_\odot$ yr$^{-1}$ which corresponds to at most 4--15\% of
the stellar mass forming within the previous 200 Myr. However this does not change the formation history derived for the massive stellar population
revealed by the MOSFIRE spectrum.

\begin{table}
\footnotesize
\centering
\caption{\bf Physical parameters of \ourobject.}
\medskip
\begin{tabular}{lcp{2cm}}
\hline
 Parameter & Value$^*$ & Comment  \\
 \hline
Photometric redshift &  $ 3.55 \pm 0.06$ & from Ref [15] \\
Spectroscopic redshift & $ 3.717 \pm 0.001$ & \\
Stellar Mass  & $(1.46-1.82)\times 10^{11}$ M$_\odot$  & From equivalent width constrained SED fit at spectroscopic redshift\\
Effective Radius & $0.49\pm 0.12$ kpc & radius enclosing half the stellar mass, from Ref [16] \\
Current star formation rate & $<0.2$ M$_\odot$ yr$^{-1}$ &   H$\beta$ emission line flux limit gives $<4$ M$_\odot$ yr$^{-1}$\\
Age  ($t_{obs}$) & $ 500-1050$ Myr & \\
Formation timescale ($t_{sf}$) & $<250$ Myr & \\
Peak star formation rate  & $>990$ M$_\odot$ yr$^{-1}$ & (and $>$ 350 M$_\odot$ yr$^{-1}$ at 95\% confidence)\\
Truncation amount $f_{drop}$ & $<10^{-4}$ &   \\
Metallicity & --- & no significant constraint \\
Extinction  & 0.4--0.6 mags  &  ($A_V$) \\
 \hline
 \vspace{0.2cm}
 \end{tabular} \\
 {\footnotesize
*Error ranges from SED fits (+ spectrum constraint) are based on 16, 84 percentiles (two-sided) and 68 percentiles (one-sided). $t_{obs}$, $t_{sf}$ and $f_{drop}$ are parameters of our
general star-formation history models and are defined in the Supplementary Information. }
  \end{table}

\begin{figure}
\begin{center}
\includegraphics[width=9cm]{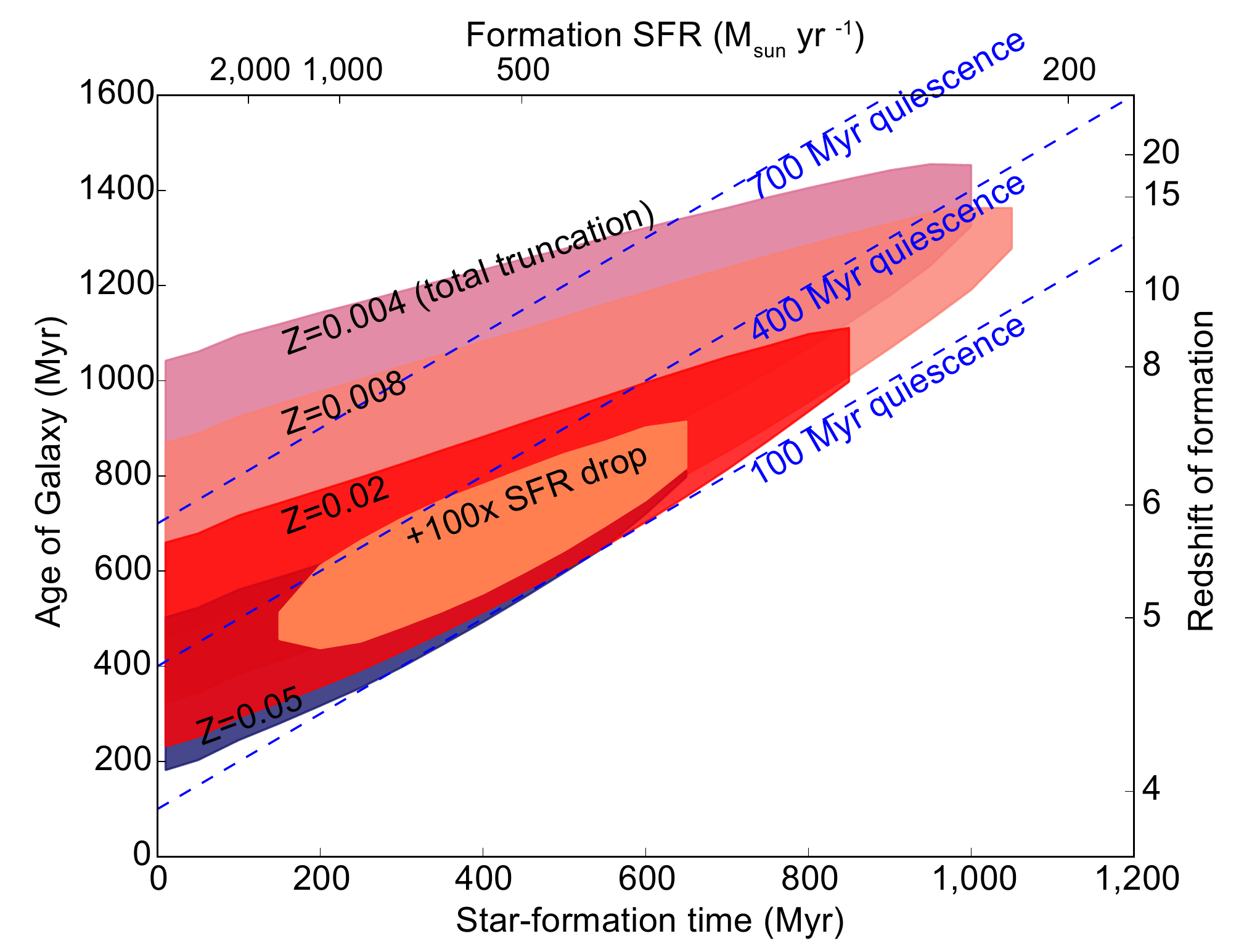} 
\end{center}
\label{fig:model}
\caption*{\small {\bf Figure 2:}  Allowed age and formation timescales of \ourobject. These constraints are
derived purely from the strength of the absorption lines in the spectrum. To accomodate
an age even as young as 200--700 Myr, the galaxy must have solar metallicity and an early
star-formation rate $>1000$ M$_\odot$ yr$^{-1}$. More extended formation times require a
much older age. The shaded areas (labelled by metallicity, with the solar metallicity
case $Z=0.02$ plotted on top for clarity) show the age and star-formation time values ($t_{obs}$ and $t_{sf}$ as defined in the Methods section) allowed within 1$\sigma$ by the equivalent width
measurement, for models where star-formation is totally shut off
during quiescence. For solar metallicity we also show the more limited range allowed if the
star-formation rate truncation is only a factor of 100. (Smaller truncation factors are
inconsistent with the line strengths). The top and right axes relate the age and star-formation
time to redshift and early star-formation rate respectively. The blue dashed lines show lines of
constant quiescence time.}
\end{figure}

There are a number of significant implications from this spectroscopic confirmation of
the existence of a quiescent galaxy population\cite{Straatman14} at $z\sim 4$ with stellar masses of $\sim 10^{11} $M$_\odot$ and a space density of $1.8\pm 0.7\times 10^{-5}$ Mpc$^{-3}$. These  are not seen in modern hydrodynamical (i.e. dark matter and baryon physics)
simulations of galaxy formation\cite{Wellons+15,MBII,Dave+16} whose volumes now approach $\sim 10^6$ Mpc$^3$. In these simulations galaxies do exist at $z\sim 4$ with similar stellar masses and abundances but they are still actively forming stars due to cosmic accretion\cite{Illustris-mergers}. They do not exist in either mode at $z>5$. The age and rapid formation time of \ourobject\ points to the formation of the the majority of the stellar mass in a single
starburst event at $z>5$, perhaps triggered by a single major merger, as opposed to a series of mergers of galaxies with different star-formation histories.
Such rapid formation is not ruled out by dynamical arguments, the compact size implies a
freefall timescale $(G\rho)^{-1/2}$ of only a few Myr.

Where are the ancestors of galaxies like \ourobject\ which must have had a star-formation rate of $\ga 1000$ M$_\odot$ yr$^{-1}$ at $z>5$? Such
galaxies are not seen in rest-frame ultraviolet censuses \cite{Smit+12,Salmon+15}. We make a plausible connection to dust-obscured star-forming galaxies\cite{Blain2002}, a handful have been spectroscopically confirmed\cite{Capak2011,ALMAz6,Ma2016}  at $z>5$ due to their intense sub-mm emission corresponding to similar star formation rates $>1000$  M$_\odot$ yr$^{-1}$. In one well-characterised case\cite{ALMAz6} at $z=6.3$ the stellar mass would have to increase five-fold to match \ourobject\
but given the object's star-formation rate this would only take 50 Myr.
Compact sub-mm galaxies at $3<z<5$ have been similarly identified\cite{Ikarashi15,Simpson15} as likely ancestors of similar compact quiescent galaxies at $z\sim 2$.
Recent deep sub-mm surveys\cite{Michalowski2016} find a space density of $3\times 10^{-6}$ Mpc$^{-3}$ for
$5<z<6$ galaxies with star-formation rates of $\ga 300$ M$_\odot$ yr$^{-1}$ (noting most redshifts were approximately estimated from the position of the sub-mm SED peak).
The ratio of space densities would
imply a short star-formation duty cycle of
$\sim$40 Myr. However none of these $z>5$ objects, in a survey volume of $\sim$ 20$\times 10^6$ Mpc$^3$, has the necessary star-formation rate $\ga 1000$ M$_\odot$ yr$^{-1}$ and the reported stellar mass
density growth contributed by the sub-mm sources is only a third of that required to make the quiescent galaxies\cite{Straatman14}.
Deeper and wider sub-mm surveys with more complete spectroscopic confirmation are required to further investigate this.
Regardless our spectroscopic confirmation  establishes that there must be
a number of such extreme early events, which
have a significant space density at $5<z<8$ (we strongly rule out redshifts $z>10$ not  yet probed  by optical/IR surveys) and are not seen in simulations\cite{Wellons+15,MBII,Dave+16}.

The space density of massive galaxies at high-redshift is an important constraint on cosmological models\cite{Steinhardt+16,Behroozi+16} as dark matter halos are growing rapidly and have to be massive and abundant enough to host them.
Our quiescent galaxy space density  at $z\sim 4$ corresponds
to a dark matter halo mass\cite{HMF}  of $\sim 3\times 10^{12}$ M$_\odot$. Thus, given the  stellar mass of \ourobject\ and a cosmic baryon fraction\cite{Planck} of 16\%, this requires 35\% of all the halo baryons to form into stars. If the galaxy formed at $z\sim 5$ then it would require 80\% of the halo baryons.  If we plot the bound on mass growth of such halos (Figure~3) it increases
very similarly with redshift to the best fit stellar mass growth from the SED implying the galaxy could not have formed significant mass prior to $z\sim 7$. We require
 a very rapid and efficient conversion of halo baryons in to stellar
mass at $5<z<6$, which is why they are not produced in current theoretical models. Conversely after $z\sim 4$ this must then quickly become  much {\it less\/} efficient. The most massive
galaxies at $z\sim 0$ have substantially lower stellar baryon fractions\cite{BG08}  of only 5--10\% (a constraint current theoretical models tune their star-formation
efficiencies to match), thus the halo must continue to grow at $z<4$ without significant further conversion of baryons to stellar mass. Rest-frame ultraviolet surveys\cite{Finkelstein+15,Steinhardt+16,Sun2016} of lower star-formation rate $z>5$ galaxies have  also found  stellar baryon fractions of $\sim$30\% at $z>5$, these are lower than needed for our ancestral population but still significantly higher than in the local Universe.

\begin{figure}
\begin{center}
\includegraphics[width=9cm]{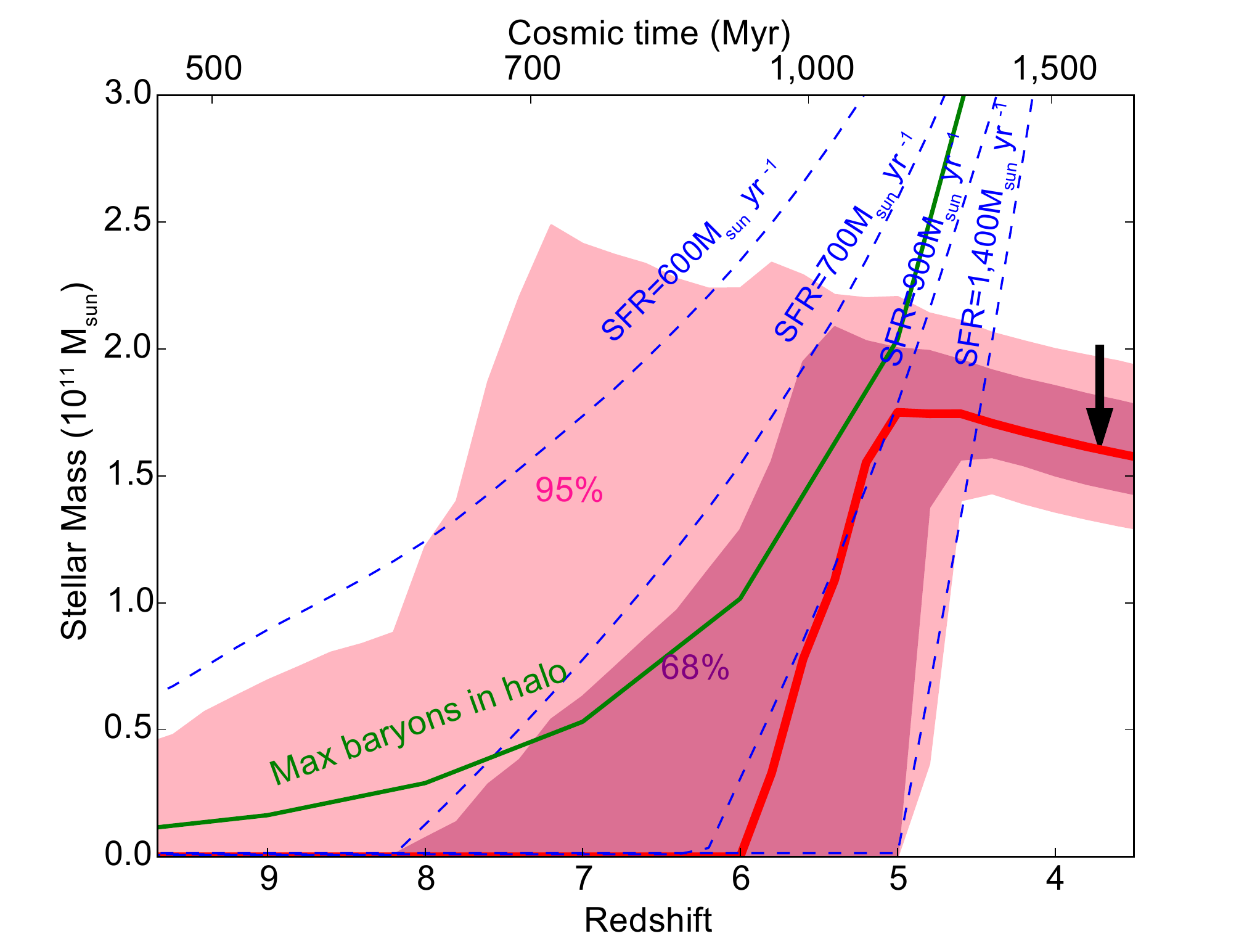}
\end{center}
\label{fig:model}
\caption*{\small {\bf Figure 3:} Stellar mass assembly history of \ourobject. These constraints are derived from the full spectroscopic and photometric data, and marginalised across the grid
of star-formation history models with rapid formation, followed by deep quiescence. The red line shows the median
stellar mass content as a function of redshift in the model grid, with the arrow denoting the epoch of observation. The colored bands show the 68\% and 95\% confidence limits on the stellar mass at each redshift. For comparison the mass growth of a set of constant
star-formation rate models, starting at different arbitrary redshifts, are shown as blue dashed lines. The green line shows
the maximum mass of baryons allowed in dark matter halos with the same number density as the $z\sim 4$ quiescent galaxies population.
\ourobject\ is robustly constrained to have been quiescent since $z\sim 5$ with a rapid earlier assembly. The slow decline in stellar mass at late times is caused by mass loss from stellar evolution.
}
\end{figure}

What is clear is that either significant revisions of the physical ingredients of galaxy formation and possibly our standard model of cold dark matter halo assembly are needed to explain the
rapid formation, and sudden and deep quenching, of massive galaxies in the very early Universe in a manner reminiscent of pre-cold dark matter
pictures of galaxy formation\cite{ELS}. Stellar mass is not a transitory phenomenon and so this observation suggests that extreme star-formation events in the early
Universe are not just rare events, they play a significant role in early mass assembly and there must be a substantial population
that will be systematically uncovered by future surveys.
We note that this is just the first spectroscopic confirmation of  a bright,  massive example of the $z\sim 4$ quiescent galaxy population. Future spectroscopic studies of the fainter quiescent population will reveal if \ourobject\ is typical. When launched, the James Webb Space Telescope will be able to get high signal-to-noise rest-frame optical spectroscopy of these galaxies and will enable detailed elemental abundances and star-formation histories to be measured for this dim red population.




\begin{addendum}
\item KG acknowledges support from Australian Research Council (ARC)
  Discovery Program (DP) grant DP130101460 and DP160102235. GGK acknowledges the support of the Australian Research Council through the award of a Future Fellowship (FT140100933). This paper is based primarily on observations taken at the W.M. Keck Observatory and the authors
 wish to recognize and acknowledge the very significant cultural role and
reverence that the summit of Mauna Kea has always had within the
indigenous Hawaiian community.  We are most fortunate to have the
opportunity to conduct observations from this mountain and we hope we
will be able to continue to do so.

\item[Author Roles] K.G. led the project, directed the observations, measured the equivalent width,
performed the star-formation history modelling analysis and wrote the paper.
C. Schreiber (C.S.)  T.N. and T.Y. reduced the MOSFIRE data. C.S. did the optimal 1D stacking and extraction of the spectra
assisted by I.L. G.K. did the mask design and assisted with the observing. Other authors assisted with
the observations and commented on the paper.

 \item[Competing Interests] The authors declare that they have no
competing financial interests.
 \item[Correspondence] Correspondence and requests for materials
should be addressed to K.G.~(email: kglazebrook@swin.edu.au).
\end{addendum}

\section*{\Large Methods}

\textbf{Data reduction and Equivalent Width Measurement}

The galaxy \ourobject\ (whose compact morphology is shown in Extended Data Figure~1) was observed using the MOSFIRE spectrograph on the Keck telescope in January and February 2016 for a combined
exposure time of 4 and 7 hours in the H and K band respectively.
The MOSFIRE 2D mask spectra were reduced and stacked using the standard data reduction pipeline with custom
modifications as described in our previous work$^{31}$. 1D spectra were extracted, respectively in the H and K bands, in individual AB pairs of 240 and 360 sec each, resulting in 59 and 71 independent spectra. These 1D spectra were flux calibrated separately and optimally stacked with inverse variance weighting. The mask included
a bright reference star to monitor seeing, transmission and the telluric absorption. Error bars were determined by using a bootstrap resampling
of the spectral stacking. We found they were 30--40\% higher than theoretical combination of poisson and readout noise as is common for
near-infrared detectors which have additional low-level systematics.

\begin{figure}
\begin{center}
\includegraphics[width=8cm]{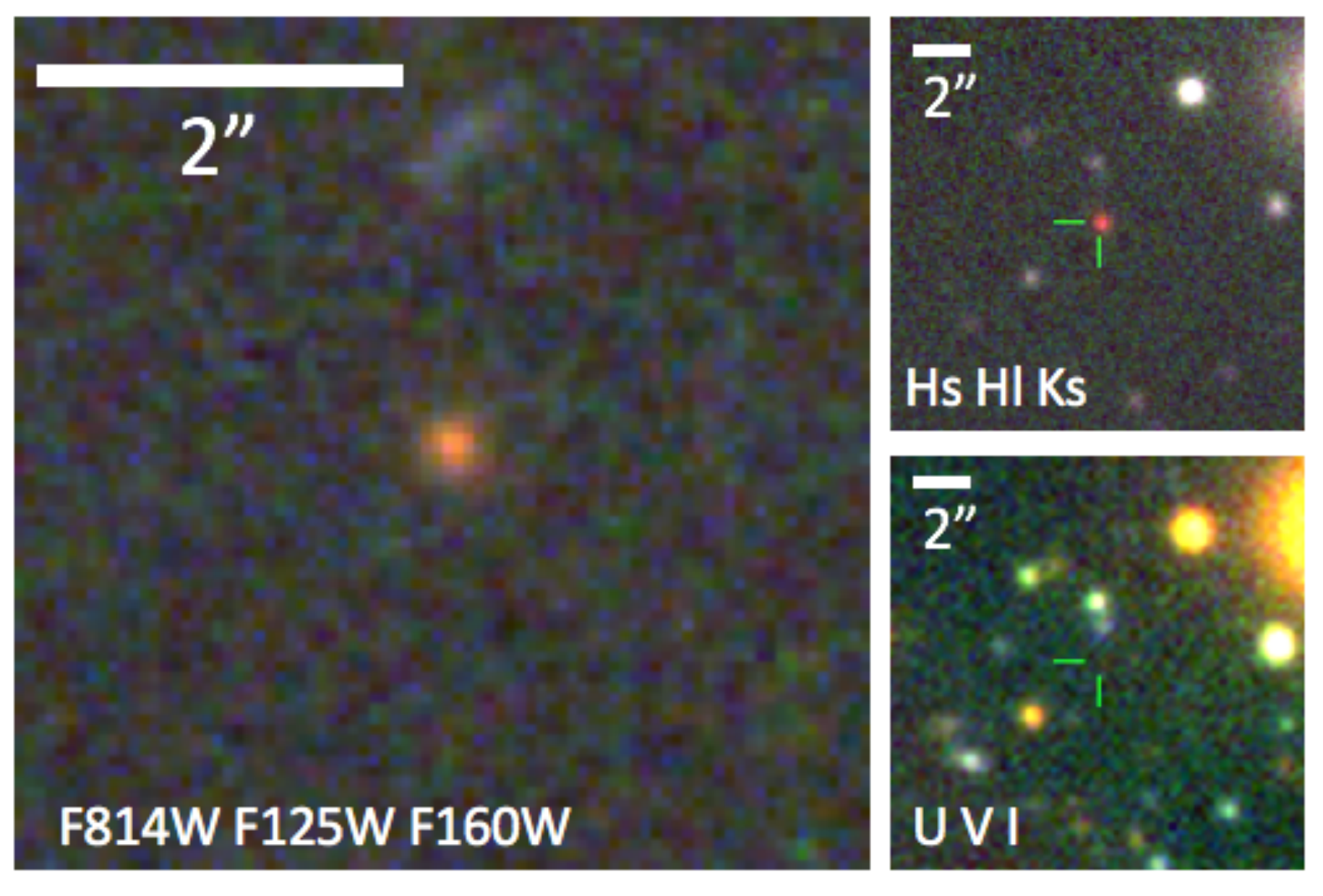}
\end{center}
\caption*{\small {\bf Extended Data Figure 1.} Imaging in the visible and near-infrared of \ourobject. The left, large, panel shows a close-up view with \textit{Hubble} (0.2 arcsec spatial resolution) and
 the right column shows wider-field and lower-resolution images from ground-based telescopes. The legend on each panel shows the mapping of
 blue, green and red colour channels to the named filters. The galaxy's flux rises strongly in the near-infrared,
 peaking at 2\micron\ wavelength
 (left and top right --- a strong red source), and is undetected in the visible below a wavelength of
 0.8\micron\ (bottom right). }
\end{figure}

Infrared spectra have to be treated carefully as there is substantial noise variation with wavelength due to  the strong sky emission
lines from OH. Further, the atmospheric absorption bands (i.e., telluric absorption) varies with time and incorrect treatment can lead to
spurious absorption lines. We did the latter in two ways: first we corrected the spectra with a master telluric correction
derived from the stacked reference star spectrum (determining the spectral type from a fit to its spectrum and photometry). Second
we used an approach where we allowed the strength of the individual telluric bands to vary in each run batch and fit for this parameter.
Both gave nearly identical final spectra, with the same galaxian absorption lines. We note that the $H\beta\gamma\delta$ lines are seen in seperate stacks of January and February
data albeit at reduced significance as expected. Further we note that the galaxian absorption lines can be seen by comparing the raw (i.e. including telluric absorption) stacked galaxy spectrum
with the pure telluric absorption spectrum.

For science analysis the spectra are binned up to a lower resolution of 19.5\AA\ per pixel, with masking of wavelengths occluded by strong sky line residuals. This gives a spectrum where the continuum is detected at median signal:noise $\simeq$ 6 in the K-band, and is shown in Figure~1 and Extended Data Figure~2. For visualisation purposes Figure~1 also shows this smoothed with a 70\AA\ gaussian filter, but all science analysis is done on the binned spectrum where the noise of each pixel is statistically independent.

We choose to measure the combined equivalent width of all three lines, to maximise the signal:noise robustly on a single measurement. We prefer here not to do full-spectrum fitting due to the
complex residuals from skylines and imperfect telluric corrections that are typically present in near-infrared spectra, and because equivalent width does not change with velocity dispersion and is
robust against exact line shape.
We measure this equivalent width by integrating with uniform weighting across all three lines to maximise the signal-to-noise ratio.
These are rest frame
bandpasses we define as 4081--4119\AA\ (H$\delta$), 4324--4374\AA\ (H$\gamma$), 4777--4900\AA\ (H$\beta$), i.e. a custom equivalent width index. The continuum level is interpolated at each wavelength from a joint linear fit to the bandpasses 4175--4236\AA, 4416--4777\AA, 4900--4950\AA, these were selected to define regions near the lines but to avoid regions affected by strong telluric corrections or potential emission lines. We measure an equivalent width of $37.6\pm 5.6$\AA\ in this index, i.e. accurate to 15\%.
Extended Data Figure~2 illustrates our equivalent width definition and measurement. To avoid any systematic bias, we apply the same procedure to the model spectra to derive their predicted equivalent widths.

\begin{figure}
\begin{center}
\includegraphics[width=9cm]{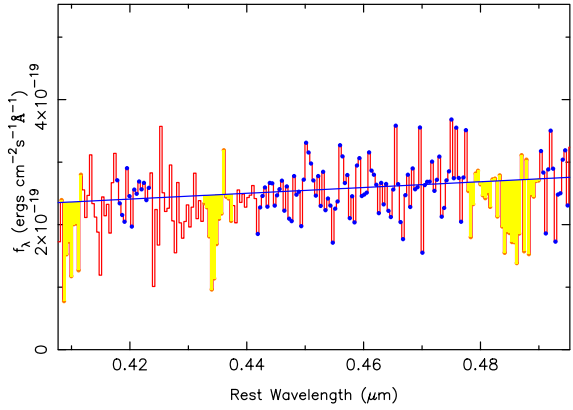}
\end{center}
\label{fig:model}
\caption*{\small{\bf Extended Data Figure 2.} Illustration of the equivalent width measurement. The red line shows the 19.5\AA\ (observed frame) spectral bins, the blue points are the clean
regions of the spectrum used for continuum fitting, the blue line is the continuum fit, and the yellow shaded areas represent the Balmer line regions summed for the equivalent width
measurement.}
\end{figure}

\textbf{Star-formation history models}

We define our star-formation model as:
\begin{enumerate}
\item The galaxy forms its stars at a constant rate $C$, for a period $t_{sf}$
\item It then truncates to form
stars at a reduced rate $C f_{drop}$ for quiescent period $t_{qu}$. For total truncation $f_{drop}=0$.
\item The galaxy is observed at an
age $t_{obs} = t_{sf} + t_{qu}$ and $t_{obs}$ must be less than 1.65 Gyr, which is the age of the Universe at $z=3.717$ (adopting a flat cosmology\cite{Planck} with $\Omega_m=0.3$, $\Omega_\lambda=0.7$, $h=0.7$).
\end{enumerate}

We used the PEGASE.2 spectral evolution of stellar population models$^{32}$ using default parameters and the Chabrier Initial Mass Function. In our model grid we allow $t_{sf}$ to vary between 10 and 1000 Myr and $f_{drop}$ to vary with a logarithmic spacing between unity and
$10^{-10}$ (i.e., essentially zero). The value of $C$ is determined from the normalisation of the stellar template when fitted to the photometry and spectrum, and the equivalent width does not depend on it. The metallicity of the models is allowed to vary between 0.004 and 0.05 (i.e 2.5 solar) and dust extinction is free to vary over $0<A_V<2$ mags. The models include emission lines of intensity proportional to the star-formation rate; strong emission lines can reduce the equivalent width or even make it go negative. For each model and time step we compute the photometric fluxes through our filters, and the equivalent width, evaluated
at the spectroscopic redshift.

\begin{figure}
\begin{center}
\includegraphics[width=9cm]{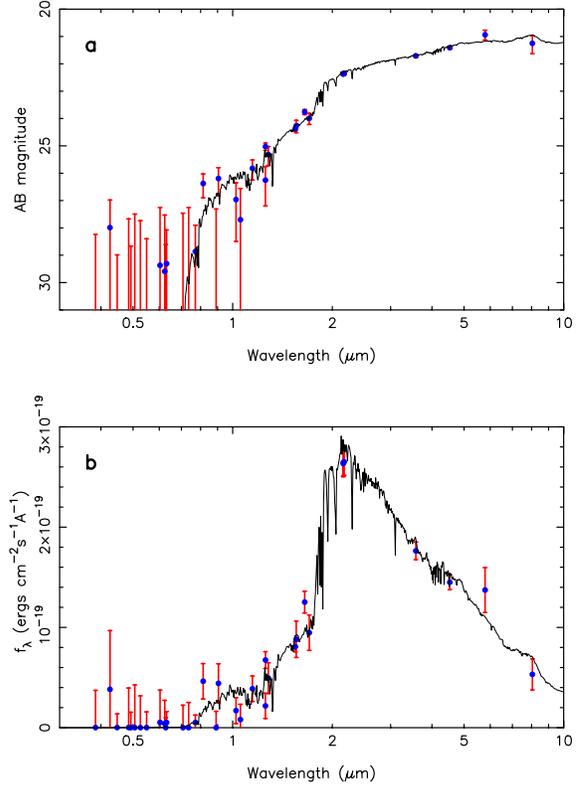}
\end{center}
\label{fig:model}
\caption*{\small {\bf Extended Data Figure 3.} Spectral energy distribution (SED) and model fits of \ourobject. The fit is performed with PEGASE.2, and is constrained by the spectroscopic redshift and equivalent widths obtained from the MOSFIRE spectrum. \textit{a}, The SED in AB magnitude (equivalent to $\log f_\nu$). \textit{b}, The same SED in $f_\lambda$. In both plots the SED is shown as a function of $\log \lambda$, and the points with error bars show the photometry measurements and their respective 1$\sigma$ uncertainty. The black line is the best fit model, which has $t_{sf} =50$ Myr and $t_{obs}=700\pm 255$ Myr (i.e.~effectively forming in a near-instantaneous burst at $z=5.8$). The age is strongly constrained by the peak at 2\micron\ in $f_{\lambda}$ and the decline redwards.
We also note the galaxy is a well detected source in the Spitzer/IRAC 3-6\micron\ images, and though the PSF there is coarse and does not allow us to resolved the galaxy, the fluxes suffer negligible flux contamination from neighboring galaxies.}
\end{figure}

The photometry$^{33}$ of \ourobject\ covers wavelengths 0.4--8.0\micron\ using a combination of ground-based optical and near-infrared imaging, Hubble Space Telescope optical/near-infrared imaging and Spitzer/IRAC space-based imaging with careful matching between bands to generate
accurate SED. We note that this source was previously attributed the ID \#13172 in earlier papers$^{15,16}$, which were based on a preliminary version of the photometric catalog.

While the photometry consists of 36 photometric data points, we only have one spectroscopic measurement --- the equivalent width. To give the latter maximum leverage,
we fit the data by minimising the $\chi^2$ across the grid with respect to the photometry, with the strong constraint that all allowed models must match the equivalent width observation within 2$\sigma$.
To estimate errors we run a gaussian Monte-Carlo perturbation of the photometric errors around the best fit model.
The best such `joint fit'  is shown in Extended Data Figure~3. The Monte-Carlo run allows us to estimate the parameter distributions. Typical fitting metallicities are 0.4--1.0 solar; super-solar models do not fit as
they only produce suitably high equivalent widths for a shorter time post-burst which then becomes inconsistent with the photometry. We find low extinctions ($A_{V}\sim 0.4$--$0.6$),
$t_{sf}\la 250$ Myr (with effectively instantaneous bursts with $t_{sf}=10$ Myr being allowed), age of $700\pm 255$ Myr and a
quiescent time $t_{qu}$ $\simeq$ 400--700 Myr.

To test the robustness of the fit we performed a jackknife test where we were deleted each point in the SED in turn. We found consistent ages and errors in all cases showing the fit is not driven by any particular point. However if we delete {\it both\/} \textit{K}-band measurements (which come from independent
telescopes) the age errors increase $\sim$ 50\%. This is expected as it is the Balmer break, peaking in the \textit{K} band, that most strongly constrains the age.

To double check these results we also used an independent code, with different underlying spectral evolution models, FAST$^{34}$ to fit the
spectrum and SED using simple SFR$(t)\propto t\,e^{-t/\tau}$ models.   We find consistent results with PEGASE.2, i.e.
($\tau=20\pm 40$ Myr) and a consistent age ($\sim 630\pm 240$ Myr). Dust extinction values are similar and metallicity is poorly constrained.
In both cases we find a best fit stellar mass of $1.7 \times 10^{11}$ M$_\odot$.

\begin{figure}
\begin{center}
\includegraphics[width=9cm,angle=0]{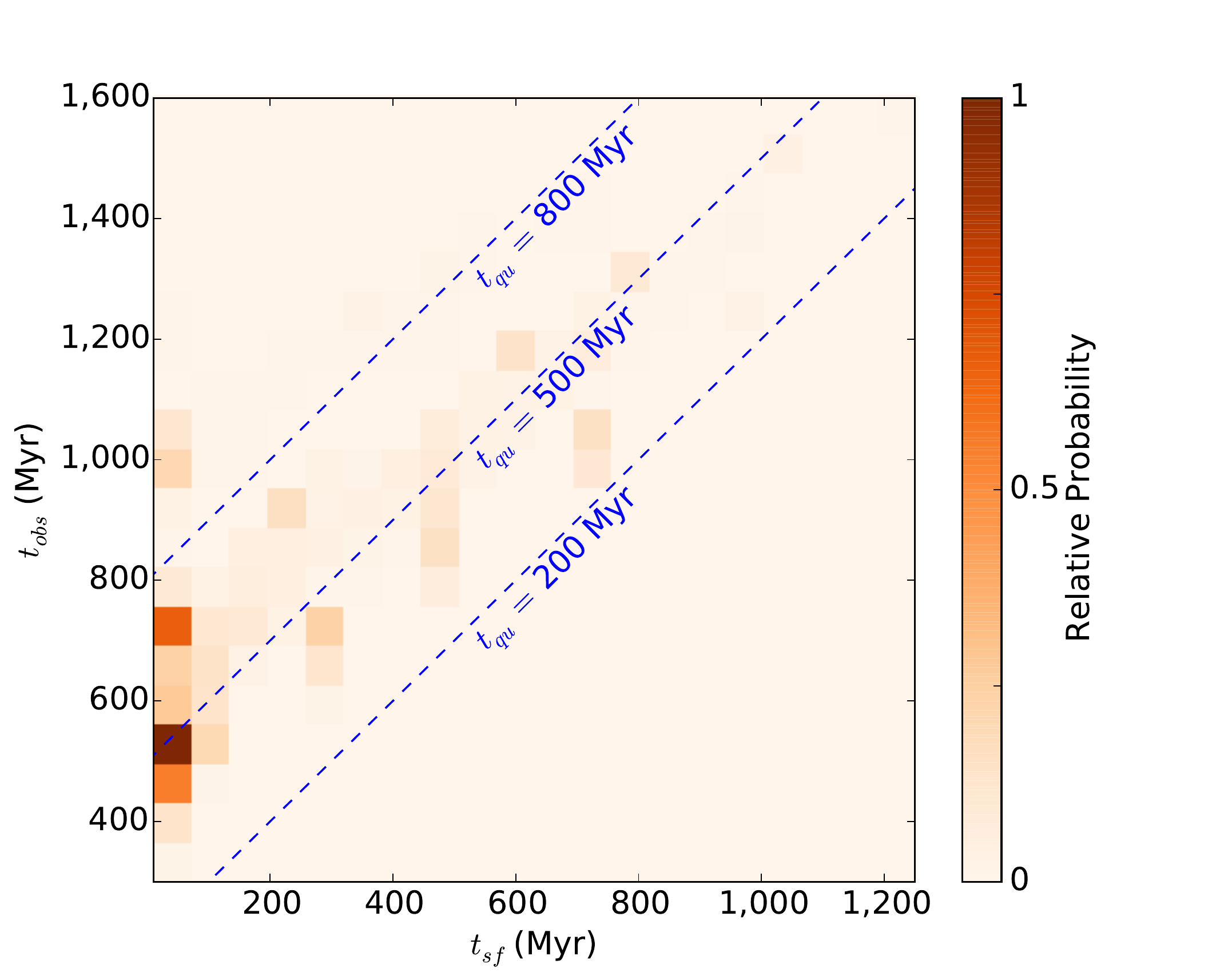}
\end{center}
\label{fig:model}
\caption*{\small{\bf Extended Data Figure 4.} Illustration of the most significant parameter degeneracy in the model
fitting, between $t_{obs}$ and $t_{sf}$. The image color scale shows the
distribution of Monte-Carlo models that fall in each bin. Very short  $t_{sf}$ is preferred though there is a tail of probability towards longer $t_{sf}$ and
 $t_{obs}$. The dashed blue lines show lines of constant quiescence time  $t_{qu}=t_{obs} - t_{sf}$. This figure shows the
 same degeneracy trend as in Figure 2, this is because the photometric age constraints have a similar sensitivity to the Balmer lines as the spectrum, via the strong
 Balmer break between the H$_{\rm long}$ (1.7\micron) and K$_{\rm s}$ (2.2\micron) bands.}
\end{figure}

The fits cover a complex parameter space of bursts lasting for different periods and observed at different ages, many of which give acceptable fits.
It is therefore not particularly illuminating to quote a single best-fitting age
and peak star-formation rate. Extended Data Figure~4 shows the parameter space of $t_{sf}$ and $t_{obs}$, which shows the principle degeneracy where
longer formation times drive older ages.
To present the constraints in the most meaningful way,
we marginalise over the fits and calculate mass assembly histories by integrating each star-formation history in the Monte-Carlo run. This integration is performed by PEGASE.2, hence accounts for recycling and other second-order processes. As mass is an integral, this is the most robust way of presenting the results, which are presented and discussed in the main text
and shown in Figure~3.

\textbf{Halo Mass calculations}

We use the online tool of [25] and compute halo mass functions for our fiducial cosmology and evaluate
the mass for which the number of halos with mass greater than this is $1.8\times 10^{-5}$ Mpc$^{-3}$. We checked several different
expressions for the `$f(\sigma)$' term in the fitting formula (Table 3 in [25]) and find the halo results do not vary significantly. For our
final calculations we adopt the $f(\sigma)$ given in their `Reed et al. 2007' table row$^{35}$, which is optimal for our redshift range.

\textbf{Long-Wavelength data}

To place limits on obscured star-formation we checked far-infrared and sub-mm surveys of the COSMOS field\cite{Straatman14,ALMAsurvey}.  The galaxy is undetected in Spitzer/MIPS and Herschel PACS/SPIRE. The 3$\sigma$ upper limits
from Herschel are 5.7, 7.2, 9.0 mJy at 250, 350, 500\micron\ respectively. ALMA detects a very faint $1.52\pm 0.25$ mJy point source at 345 Ghz a distance $0.5\pm 0.1$ arcsec
from the optical centre. There is no ALMA emission from the optical location.
The ALMA astrometry is aligned with accurate radio data from the Very Large Array in this field and a cross-check with our optical/near-IR catalogs show that any systematic errors
are $<0.05$ arcsec. For a source at $z=3.7$ the ALMA and Herschel fluxes would be consistent with maximum obscured star-formation rates $\lesssim$ 50--200 M$_\odot$ yr$^{-1}$ and an associated dust
temperature of $<$35K.  Further observations and analysis are in progress to determine if the ALMA source is at the same redshift, whether it is powered by star-formation, and 
 if is a companion galaxy or part of the same system.

\begin{addendum}
\item[Code availability] The spectrophotometric codes (PEGASE.2 and FAST) are publicly available. The  high-level {\it Perl Data Language} scripts which use PEGASE.2 to model our custom star-formation histories are available on request, noting this is a standard technique and is easily reproduced.
\item[Data availability] The spectrum (Figure 1) and best fit mass-assembly histories (Figure 3) of \ourobject\ are made available along with this manuscript as Source Data. The other data that support the plots within this paper and other findings of this study are available from the corresponding author upon reasonable request.
\end{addendum}

\end{document}